# Ensuring Outcome-Based Curriculum Coherence through Systematic *CLO–PLO* Alignment and Feedback Loops

**Moncef Derouich**⋆

Astronomy and Space Science Department, Faculty of Science, King Abdulaziz University, PO Box 80203, Jeddah 21589, Saudi Arabia

**Abstract.** This study proposes a quantitative framework to enhance curriculum coherence through the systematic alignment of Course Learning Outcomes (*CLO*s) and Program Learning Outcomes (*PLO*s), contributing directly to the continuous improvement of outcome-based education. Grounded in widely recognized accreditation standards such as ABET and NCAAA, the model introduces mathematical tools that map exercises, assessment questions, teaching units (TUs), and student assessment components (SACs) to *CLO*s and *PLO*s. This dual-layer approach—combining micro-level analysis of assessment elements with macro-level curriculum evaluation—enables detailed tracking of learning outcomes and helps identify misalignments between instructional delivery, assessment strategies, and program objectives. The framework incorporates alignment matrices, weighted relationships, and practical indicators to quantify coherence and evaluate the overall performance of a course or program. Application of this model reveals gaps in outcome coverage and underscores the importance of realignment, especially in cases where specific *PLO*s are underrepresented or *CLO*s are not adequately supported by assessments. The proposed model is practical, adaptable, and scalable, making it suitable for academic programs in science, technology, engineering, and mathematics (STEM). Its systematic structure supports institutions in implementing evidence-based curriculum enhancements and provides a reliable mechanism for aligning teaching practices with desired learning outcomes. Ultimately, this framework offers a valuable tool for closing the feedback loop between instructional design, assessment execution, and learning outcomes, thereby promoting greater transparency, accountability, and educational effectiveness. Institutions that adopt this model can expect to strengthen their quality assurance processes and ensure that students graduate with the competencies required by academic standards and professional expectations.

⋆ Corresponding author: derouichmoncef@gmail.com & aldarwish@kau.edu.sa

# 1 Introduction

Constructive alignment is a cornerstone of curriculum design, linking teaching, learning, and assessment within a coherent framework (Biggs & Tang, 2011). In professional fields such as engineering, effective alignment ensures that instructional activities genuinely support intended outcomes (Felder & Brent, 2003), and diversified, outcomes-focused assessment strengthens this connection (Craddock & Mathias, 2009). Despite this consensus, implementation remains difficult. Differences in how instructors teach and how students prefer to learn—e.g., an emphasis on abstract theory when learners benefit from concrete, applied problem-solving—can depress engagement (Felder & Silverman, 1988). More critically, assessment practices often emphasize tasks that are straightforward to grade—for example, multiple-choice items, short factual answers, or single-solution exercises—because they are efficient and consistent, especially at scale; yet these formats chiefly capture lower-order skills (recall, routine procedures) and can neglect higher-order outcomes such as analysis, synthesis, and authentic problem-solving that programs explicitly value (Reeves, 2006). Institutions also face persistent challenges embedding Outcome-Based Education (OBE) within accreditation constraints (Patil & Codner, 2007). Foundational curriculum models have long emphasized that aims, instruction, and assessment must operate in concert for learning to be effective (Harden, 2001; Tyler, 1949).

Recent work spans faculty evaluation, sector-level improvement, instrument design, institutional OBE indices, and accreditation-driven attainment procedures (Agrawal & Sharma, 2025; Goba et al., 2024; Bernold & Díaz-Michell, 2022; Lu & Hao, 2022; Beena & Suresh, 2021). Relative to these strands, the present study introduces a quantitative, compositional alignment framework with two defining features. First, at the micro level, we specify a weighted mapping from assessment items (questions/exercises) $\rightarrow$ Course Learning Outcomes ($CLO$s) $\rightarrow$ Program Learning Outcomes ($PLO$s), yielding closed-form expressions that make each item's contribution explicit and auditable (Craddock & Mathias, 2009; Mak & Frezza, 2006). Second, at the macro level, we quantify intended versus delivered emphasis in Teaching Units (TUs) and Student Assessment Components (SACs) via Intended Teaching Units (ITU) and Instructional Support and Assessment Coverage (ISAC), with acceptance bands to separate normal variation from patterns warranting action (Felder & Brent, 2003; NCAAA, 2021). We then relate these indicators to student-performance evidence to close the feedback loop and guide targeted adjustments (Reeves, 2006; Wanous, Procter, & Thomas, 2009). The formulation also clarifies program distribution: a course may, by design, contribute minimally to selected $PLO$s if coverage is secured elsewhere (González & Wagenaar, 2008; ENQA, 2015; NCAAA, 2021).

Positioning and distinctiveness relative to existing frameworks. The proposed framework differs from prior contributions in the unit of analysis and the methodological design. Agrawal & Sharma (2025) prototype a system for faculty performance evaluation; our focus is curriculum-level alignment across courses and programs. Goba et al. (2024) synthesize quality-improvement strategies at a policy/sector level; we operate at design/execution level with quantitative $CLO$–$PLO$ mapping. Bernold & Díaz-Michell (2022) offer a multi-dimensional assessment instrument centered on constructivist interaction; our model is compositional, tracing item-to-$PLO$ contributions and TU/SAC aggregation. Lu & Hao (2022) develop institutional evaluation indices using multi-criteria methods; in contrast, we analyze curricular coherence through explicit mappings and tolerance bands. Beena & Suresh (2021) present accreditation-aligned CO–PO/PSO correlation and attainment averages; we move beyond correlations to an auditable, equation-based mapping that (i) decomposes item-level evidence to *PLO*s, (ii) integrates TU/SAC signals via metrics that evaluate the alignment of TU and SAC, and (iii) generates actionable flags when acceptance bands are exceeded. In short, whereas prior work emphasizes faculty, policy, instrument, or institutional index perspectives, our contribution is a generalized, quantitative curriculum-alignment framework that links assessment items and teaching units to program outcomes, revealing precise alignment gaps and enabling principled corrections.

Astronomy curricula are inherently interdisciplinary, integrating core areas of physics—mechanics, electromagnetism, and quantum or statistical physics—with a strong mathematical foundation in calculus, linear algebra, and numerical methods. These are complemented by observational and experimental components, including laboratory work, instrumentation, and data analysis. The proposed framework is well suited to managing and evaluating such complex structures.

By enabling transparent mapping of item-level evidence from physics- and math-intensive courses to astronomy-specific *CLO*s and *PLO*s, the framework supports a rigorous audit of curricular alignment. It helps evaluate the balance between theoretical instruction and observational or assessment-based components, ensuring that no essential outcomes are overlooked. Importantly, it also captures—by design—instances where certain astronomy *PLO*s are not addressed in specific courses because they are intentionally covered in companion laboratory, methods, or capstone experiences. This structured visibility allows departments to demonstrate a coherent progression from foundational skills to astronomy-specific competencies, reducing ambiguity and supporting program-level coherence.

The aim of this work is to evaluate whether the proposed framework ensures comprehensive coverage between *CLO*s and *PLO*s while identifying actionable misalignments. These misalignments are quantified through suitable parameters, and defined acceptance bands, with the overarching goal of strengthening continuous improvement in curriculum alignment and delivery.

The paper is organized into five sections. Section II introduces the foundational approach based on the alignment of exercises, assessment questions, *CLO*s, and *PLO*s, including a worked example and the combined expression from questions to *PLO*s. Section III develops a broader audit using TUs, SACs, indicators that measure the alignment of TUs and SACs, and acceptance bands. Section IV presents key observations and recommendations (e.g., uncovered *PLO*s, overlap, focused mapping). Section V details closing the loop through re-alignment of TUs and SACs with *CLO*s and *PLO*s, embedding mechanisms for institutionalizing continuous improvement. Section VI concludes.

## 2 Approach Based on Alignment of Exercises, Questions, *CLO*s, and *PLO*s

### 2.1 Alignment of *PLO*s with *CLO*s

The alignment of *PLO*s with *CLO*s is designed to ensure students develop comprehensive knowledge, skills, and competencies throughout their academic journey. This alignment integrates theoretical foundations, practical applications, and interdisciplinary problem-solving strategies, all tailored to meet the objectives of the program. The process includes the mapping of *PLO*s to specific courses and assessing their contribution to student achievement, an approach widely used in accreditation-driven case studies (e.g., Beena & Suresh, 2021).

The alignment of *PLO*s with *CLO*s is performed according to the following considerations:

- Structured Mapping:
  - Each *PLO* is mapped to multiple courses, ensuring comprehensive coverage across different stages of the curriculum.
  - For example, foundational courses address introductory-level knowledge, while advanced courses tackle specialized skills and applications. Prior studies confirm that such structured matrices are essential for curriculum quality audits (e.g. Lu & Hao, 2022).
- Iterative Assessment:
  - *PLO* attainment is evaluated through both direct (e.g., exams, assignments) and indirect methods (e.g., surveys, reflective feedback).
  - Faculty analyse course-level data to identify gaps and recommend adjustments in course content or assessment strategies. Faculty-centered performance systems have been proposed as one mechanism to improve this loop (e.g., Agrawal & Sharma, 2025), though our focus here is curriculum alignment rather than instructor evaluation.
- Continuous Improvement:
  - The alignment meetings ensure coherence by systematically reviewing *CLO–PLO* matrices, comparing intended outcomes with assessment data, and documenting any gaps. This structured process guarantees that adjustments are based on evidence rather than informal impressions. At the institutional and sector level, quality-improvement strategies (Goba et al., 2024) and multi-dimensional assessment instruments (Bernold & Díaz-Michell, 2022) have been proposed. Our contribution complements these by working at the compositional level—examining how individual questions, exercises, and teaching units aggregate into *CLO*s and *PLO*s, and by developing indicators that quantify their alignment.
  - The role of faculty is to analyze course-level results and recommend revisions to *CLO*s, while advisory board members contribute external and professional perspectives to validate relevance. Together, these contributions ensure that alignment is both academically rigorous and responsive to industry expectations.

While the general framework provides a conceptual foundation for aligning *PLO*s with *CLO*s, a more precise and quantitative approach is necessary to ensure that the alignment is measurable, consistent, and effective. To achieve this, we introduce a mathematical model that formalizes the relationships between individual assessment components (e.g., exercises, questions) and the broader learning outcomes (*CLO*s and *PLO*s). Relevant mapping and quantitative outcome assessment work has been done. For example, Zayani et al. (2024) introduced an automated framework for assessing *CLO*-to-*PLO* relationships using weighted aggregation across course sections. Mendoza et al. (2022) analyze curriculum mapping at macro, meso, and micro levels. Mustaffa et al. (2019) provide quantitative measurement of *CLO* achievement in large classes using multiple metrics. These works inform but do not decompose down to the level of individual questions or teaching units in the way our weighted mapping model does.

## 2.2 Equations Linking Exercises, Questions, *CLO*s, and *PLO*s

### 2.2.1 Definitions

Exercises here designate problem-solving tasks, assignments, projects, lab work, presentations, and exams, each designed to evaluate students' understanding and skills. This section outlines the mathematical relationships between question-level outcomes ($Q_{li}$, the $i$-th question in the $l$-th exercise), *CLO*s ($CLO_1$, $CLO_2$, $CLO_3$, ...), and *PLO*s ($PLO_1$, $PLO_2$, $PLO_3$, ...). These relationships are modeled through weighted contributions. The Exercises-*CLO* Mapping Matrix aligns structured learning activities with specific *CLO*s to ensure comprehensive assessment. Some exercises, like Exercise 2, may be composed of multiple questions ($Q_{21}$, $Q_{22}$, $Q_{23}$, $Q_{24}$, $Q_{25}$, $Q_{26}$), each targeting different *CLO*s. For example, $Q_{22}$ assesses $CLO_3$, while $Q_{11}$ maps to $CLO_1$, and $Q_{22}$, $Q_{23}$, $Q_{24}$, $Q_{25}$, $Q_{26}$ serve to evaluate $CLO_3$. This structured mapping ensures that exercises contribute effectively to measuring and achieving the intended learning outcomes.

*Table 1: Exercises-CLO Mapping Matrix*

| Exercise | Questions | Mapped CLOs |
|---|---|---|
| Exercise 1 | $Q_{11}, Q_{12}, Q_{13}$ | $CLO_1\ (Q_{11}), CLO_2\ (Q_{12}, Q_{13})$ |
| Exercise 2 | $Q_{21}, Q_{22}, Q_{23}, Q_{24}, Q_{25}, Q_{26}$ | $CLO_1\ (Q_{21}), CLO_3\ (Q_{22}, Q_{23}, Q_{24}, Q_{25}, Q_{26}), CLO_5$ $(Q_{26})$ |

*Table 2: CLO-PLO mapping matrix*

| $PLO_1$ | $PLO_2$ | $PLO_3$ | $PLO_4$ | $PLO_5$ | $PLO_6$ |
|---|---|---|---|---|---|
| $CLO_1$ $CLO_2$ $CLO_5$ | $CLO_2$ $CLO_4$ | | | $CLO_6$ | |

### 2.2.2 Mathematical approach for mapping Questions to *CLO*s

Each question $Q_{li}$ in the exercises is designed to assess specific *CLO*s, as defined by the course objectives. For example, in Table 1, Exercise 1 evaluates $CLO_1$ and $CLO_2$. Exercise 2 evaluates:

- $CLO_1$ (through $Q_{21}$),
- $CLO_3$ (through $Q_{22}, Q_{23}, Q_{24}, Q_{25}, Q_{26}$),
- $CLO_5$ (through $Q_{26}$)

The mapping can be described using weighted contributions:

$$CLO_j = \sum_{l=1}^{m} \sum_{i=1}^{n} w_{lij}\ Q_{li} \qquad (1),$$

where:
$CLO_j$: The $j$-th Course Learning Outcome.

$Q_{li}$: The $i$-th question in the $l$-th exercise.

$w_{lij}$: a weight factor that represents the contribution of question $Q_{li}$ to $CLO_j$.

- $m$: Total number of exercises.

- $n$: Total number of questions per exercise.

### 2.2.3 Mapping *CLO*s to *PLO*s

Following standard curriculum-mapping practice, we assume each *PLO* is supported by multiple *CLOs* with unequal emphasis; that emphasis is represented by non-negative weights that are normalized to sum to 1 across the contributing *CLOs* (Harden, 2001; NCAAA, 2021; Beena & Suresh, 2021; Lu & Hao, 2022; González & Wagenaar, 2008). Each *CLO* contributes to one or more *PLOs* according to this equation:

$$PLO_k = \sum_{g=1}^{p} v_{gk} \ CLO_g \qquad (2),$$

where:

- $PLO_k$: the $k$-th Program Learning Outcome.

- $CLO_g$: the $g$-th Course Learning Outcome.

- $v_{gk}$: weight of $CLO_g$ contributing to $PLO_k$

- $p$: the total number of $CLOs$.

### 2.2.4 Combined Relationship: Questions to *PLO*s

By substituting the equation for *CLOs* into the equation for *PLOs*, the relationship between questions and *PLOs* can be expressed as:

$$PLO_k = \sum_{g=1}^{p} v_{gk} \left( \sum_{l=1}^{m} \sum_{i=1}^{n} w_{lig} \ Q_{li} \right)$$

$$\Rightarrow PLO_k = \sum_{g=1}^{p} \sum_{l=1}^{m} \sum_{i=1}^{n} v_{gk} \ w_{lig} \ Q_{li} \qquad (3)$$

-Each question $Q_{li}$ contributes to $CLO_g$ based on weights $w_{lig}$.

-Each $CLO_g$ contributes to $PLO_k$ based on weights $v_{gk}$.

-Combined weights $v_{gk} w_{lig}$ determine the relative contribution of each question $Q_{li}$ to each $PLO_k$. For instance, if $v_{gk} w_{lig} = 0$, it means $Q_{li}$ does not contribute to $PLO_k$. The $v_{gk} w_{lig}$ weights aggregate the contribution of each question to the $PLO_k$, allowing a quantitative measure of how assessments impact program objectives.

As an example of application of the Mathematical formalism, using Table 1, through Equation (1) one can identify how individual questions contribute to specific *CLO*s:

$CLO_1 = w_{111} \ Q_{11} + w_{211} \ Q_{21}$

$CLO_2 = w_{122} \ Q_{12} + w_{132} \ Q_{13}$

$CLO_3 = w_{223} \ Q_{22} + w_{233} \ Q_{23} + w_{243} \ Q_{24} + w_{253} \ Q_{25} + w_{263} \ Q_{26}$

$CLO_4 = 0$

$CLO_5 = w_{265}\, Q_{26}$

$CLO_6 = 0$

Using Table 2 and Equation (2), the combined relationships relating questions to PLOs equations can be written numerically as:

$PLO_1 = v_{11}\, CLO_1 + v_{31}\, CLO_3 + v_{51}\, CLO_5$
$\quad = v_{11}\, w_{111}\, Q_{11} + v_{11}\, w_{211}\, Q_{21} + v_{31}\, w_{223}\, Q_{22} + v_{31}\, w_{233}\, Q_{23} + v_{31}\, w_{243}\, Q_{24}$
$\quad + v_{31}\, w_{253}\, Q_{25} + v_{31}\, w_{263}\, Q_{26} + v_{51}\, w_{265}\, Q_{26}$

$PLO_2 = v_{22}\, CLO_2 + v_{42}\, CLO_4 = v_{22}\, w_{122}\, Q_{12} + v_{22}\, w_{132}\, Q_{13}$

$PLO_3 = 0$

$PLO_4 = 0$

$PLO_5 = v_{65}\, CLO_6 = 0$

$PLO_6 = 0$

In concrete terms, by design, the teacher can propose an exam where the questions are related to the *CLO*s by matrix of Table 3. Thus:

$CLO_1 = w_{111}\, Q_{11} + w_{211}\, Q_{21} = 0.5\, Q_{11} + 0.5\, Q_{21}$

$CLO_2 = 0.3\, Q_{12} + 0.7\, Q_{13}$

$CLO_3 = 0.2\, Q_{22} + 0.2\, Q_{23} + 0.2\, Q_{24} + 0.2\, Q_{25} + 0.2\, Q_{26}$

$CLO_4 = 0$

$CLO_5 = 0.8\, Q_{26}$

$CLO_6 = 0$

*Table3: Exercises-CLO Mapping Matrix with Weights $w_{lij}$*

| $Q_{li}/CLO_j$ | $CLO_1$ | $CLO_2$ | $CLO_3$ | $CLO_4$ | $CLO_5$ | $CLO_6$ |
|---|---|---|---|---|---|---|
| $Q_{11}$ | 0.5 | -- | -- | -- | -- | -- |
| $Q_{12}$ | -- | 0.3 | -- | -- | -- | -- |
| $Q_{13}$ | -- | 0.7 | -- | -- | -- | -- |
| $Q_{21}$ | 0.5 | -- | -- | -- | -- | -- |
| $Q_{22}$ | -- | -- | 0.2 | -- | -- | -- |
| $Q_{23}$ | -- | -- | 0.2 | -- | -- | -- |
| $Q_{24}$ | -- | -- | 0.2 | -- | -- | -- |
| $Q_{25}$ | -- | -- | 0.2 | -- | -- | -- |
| $Q_{26}$ | -- | -- | 0.2 | -- | 0.8 | -- |

*Table 4: CLO-PLO Mapping Matrix with Weights $v_{gk}$*

| $CLO_g/PLO_k$ | $PLO_1$ | $PLO_2$ | $PLO_3$ | $PLO_4$ | $PLO_5$ | $PLO_6$ |
|---|---|---|---|---|---|---|
| $CLO_1$ | 0.4 | -- | -- | -- | -- | -- |
| $CLO_2$ | -- | 0.6 | -- | -- | -- | -- |
| $CLO_3$ | 0.4 | -- | -- | -- | -- | -- |
| $CLO_4$ | -- | 0.4 | -- | -- | -- | -- |
| $CLO_5$ | 0.2 | -- | -- | -- | -- | -- |
| $CLO_6$ | -- | -- | -- | -- | 1.0 | -- |

Therefore, using the *CLO-PLO* mapping matrix shown in Table 4, we obtain:

$PLO_1 = 0.4\ CLO_1 + 0.4\ CLO_3 + 0.2\ CLO_5$

$PLO_2 = 0.6\ CLO_2 + 0.4\ CLO_4 = 0.6\ CLO_2$ since $CLO_4=0$ (see above)

$PLO_3 = 0$

$PLO_4 = 0$

$PLO_5 = CLO_6$

$PLO_6 = 0$

Thus, one can determine the relative contribution of each question $Q_{li}$ to each $PLO_k$ by substituting the *CLO*s with their expressions in terms of the $Q_{li}$, following the principle of constructive alignment (e.g. Biggs, 1996; Harden, 2001). $PLO_3$, $PLO_4$, and $PLO_6$ evaluate to zero because no *CLO* in this course maps to them. A zero at the course level indicates that the *PLO* is outside the intended scope of this course, not that it is neglected by the program. Even if this course doesn't address certain *PLOs*, the program as a whole still does—because other courses cover them.

## 2.3    Observations and Recommendations

### 2.3.1    Observations

a) **Gaps in Alignment**:
   o $PLO_3$, $PLO_4$, and $PLO_6$ are not mapped to any *CLO*, which could mean certain program objectives are not addressed through the course design (e.g. Harden, 2001).
   o This leaves room for improvement to ensure comprehensive curriculum coverage.
b) **Overlapping *CLO*s for $PLO_1$**:
   o $PLO_1$ is mapped to three *CLO*s ($CLO_1$, $CLO_2$, $CLO_5$), suggesting a strong emphasis on the outcomes tied to this program-level objective.
   o Consider redistributing *CLO*s to reduce redundancy and focus on addressing gaps.
c) **Focused Mapping**:
   o $PLO_5$ is mapped to only one *CLO* ($CLO_6$), which may indicate targeted coverage but could also mean insufficient breadth for this outcome. Note that $CLO_6$ is not covered by any exercise ($PLO_6$ =0) (e.g., Biggs, 1996).

### 2.3.2 Recommendations

a) **Address Gaps**:
  - o Some courses include a *CLO-PLO* mapping matrix that may not align with certain *PLO*s. For instance, as shown in Table 1, one course does not address $PLO_3$, $PLO_4$, and $PLO_6$. In these cases, it is necessary for other courses in the study plan to cover the missing *PLO*s (Harden, 2001; NCAAA, 2023).
  - o The study plan (i.e., the curriculum) should be balanced to ensure that all *PLO*s are achieved and that courses complement one another. Therefore, it is essential to evaluate course objectives to confirm that every program outcome is adequately addressed.
b) **Curriculum Review**:
  - o Perform a detailed review of the curriculum to identify opportunities for adding or refining *CLOs* to ensure alignment with the full set of *PLOs* (Biggs, 1996; Harden, 2001).

## 3   More general Approach

## 3.1   Approach adequacy

The Exercises-Questions-*CLO*s-*PLO*s approach presented in the previous section is more suited for assessment design than for curriculum design. It focuses on ensuring that individual questions align with learning outcomes, providing micro-level insights into how specific exercises and questions contribute to *CLO*s and *PLO*s (Harden 2001, Downing 2006). While this granularity is valuable for detailed analysis, it may overlook the broader curriculum goals and the holistic impact of teaching and assessment methods on student learning (Biggs 1996, Prideaux 2003). By concentrating on specific exercises and questions, this approach risks missing the bigger picture of how teaching and assessment strategies collectively contribute to achieving learning outcomes (Biggs 1996, Wiggins & McTighe 2005).

To address this limitation, a more comprehensive approach can be employed to infer *CLO* and *PLO* attainment rates by incorporating both delivered TUs and SACs into the evaluation process. This approach links TUs and SACs to *CLO*s and *PLO*s, providing a more holistic view of curriculum alignment and effectiveness (Harden, 2001; Prideaux, 2003; ABET, 2025–2026; NCAAA, 2023).

TUs should align with the *CLO*s and reflect the intended emphasis of each outcome. The emphasis of *CLO*s in TUs can vary depending on the teaching approach, student needs, and instructional adjustments. Misalignment of TUs with *CLO*s over multiple course deliveries suggests the need for curriculum revision. For each course, SACs should cover specific *CLO*s and reflect their intended importance. The emphasis of *CLO*s in SACs can be quantified to assess their alignment. A misalignment between TUs and SACs (i.e., if a *CLO* is not assessed adequately) indicates a potential gap in curriculum delivery (Harden, 2001; Prideaux 2003; ABET,  2025–2026; NCAAA,  2023).

This approach is flexible and can be applied to a wide range of teaching and assessment methods, regardless of the specific exercises or questions used. It is adaptable to different courses, disciplines, and educational contexts, making it more general and widely applicable. This approach emphasizes the teaching and assessment processes as a whole, rather than individual components like exercises or questions. It aligns with modern educational practices that focus on outcome-based education (OBE), where the emphasis is on how teaching and assessment methods collectively contribute to achieving learning outcomes (Spady, 1994; ABET, 2025–2026; NCAAA, 2023).

## 3.2 TUs and Student Assessment Components (SACs)

### 3.2.1 Teaching Units (TUs)

- **Definition**: TUs encompass all teaching and learning activities, such as lectures, tutorials, labs, projects, and presentations. This approach considers the entire teaching process, not just specific exercises or questions.
- **Purpose**: It allows for a holistic view of how teaching methods contribute to achieving *CLO*s and *PLO*s. By analyzing the time and effort spent on each topic, institutions can ensure that the intended emphasis of each *CLO* is reflected in the delivered curriculum.
- **Alignment**: TUs should align with the *CLO*s and reflect the intended emphasis of each outcome. Misalignment of TUs with *CLO*s over multiple course deliveries suggests the need for curriculum revision.

### 3.2.2 Student Assessment Components (SACs)

- **Definition**: SACs include all assessment methods, such as exams, quizzes, homework, lab reports, and group projects. This approach considers the entire assessment process, not just individual questions or exercises (see Biggs, 1996; Harden, 2001; Prideaux, 2003).
- **Purpose**: It provides a comprehensive view of how assessments measure student achievement of *CLO*s and *PLO*s. By ensuring that SACs cover all *CLO*s and reflect their intended importance, institutions can better evaluate the effectiveness of their curriculum.
- **Alignment**: SACs should align with the intended emphasis of *CLO*s. A misalignment between TUs and SACs (e.g., if a *CLO* is not assessed adequately) indicates a potential gap in curriculum delivery (see Biggs, 1996; Wiggins & McTighe, 2005; Prideaux, 2003).

## 3.3 Mathematical formulation

### 3.3.1 Estimation of emphasis of CLOs in delivered TUs

A mathematical method, similar to the one presented in the previous section, can be used to develop indicators that measure the alignment of TU and SAC with *CLOs* and *PLOs*. These indicators include:

- Emphasis of *CLO* and *PLO* in delivered TU: $I_{TU\text{-}CLO}$ and $I_{TU\text{-}PLO}$
- Emphasis of *CLO* and *PLO* in SAC: $I_{SAC\text{-}CO}$ and $I_{SAC\text{-}PLO}$
- Alignment of SAC with TU: $I_{SAC\text{-}CO}/I_{TU\text{-}CO}$ and $I_{SAC\text{-}PLO}/I_{TU\text{-}PLO}$

These indicators offer a systematic and rational basis for aligning TU and SAC with *CLO* and *PLO*. Additionally, they support the revision of *CLO* and the re-alignment of *CLO* with *PLO* as needed. The estimation process can be automated using a spreadsheet, which facilitates comparison of results against user-defined limits and generates a comprehensive summary report, including tables and figures.

Table 5 outlines the planned weight for each *CLO* as designed by the course architect. For example, $CLO_1$ was assigned a weight of 4 (28.6% of the overall emphasis). The table then shows how many lectures and tutorials were actually dedicated to each outcome, converting those session counts into an effective delivered weight. A weighted system was applied—lectures contributing 0.8 and tutorials 0.2—to convert the number of sessions into a percentage of the total Teaching Unit (TU). Table 5 employs a four-point scale, where 1 indicates minimal emphasis and 4 indicates maximum emphasis, to map the Teaching Unit to each *CLO*.

For instance, for $CLO_1$, the calculation is as follows:

- Lectures: 26.7 % of total lecture time × 0.8 = 21.36 %
- Tutorials: 25 % of total tutorial time × 0.2 = 5 %
- Total Delivered Emphasis: 21.36 % + 5 % = 26.36 %, equivalent to a weight: 26.36/100 x 15 = 3.95

This table is critical for identifying discrepancies between the designed emphasis and what was actually delivered. It highlights where the course might have underemphasized or overemphasized certain outcomes relative to the original plan. The Delivered emphasis of *CLO* in TU calculated in Table 5 by using the relation:

$$Delivered \; Weight = \left( \frac{Combined \; Delivered \; Emphasis}{100} \right) \times 15$$

*Table 5: Estimation of emphasis of CLO in delivered TU. To add lectures and tutorials, the weight assigned to lectures was 0.8 and the weight assigned to tutorials was 0.2. We use a four-point scale--where 1 signifies low emphasis and 4 represents maximum emphasis--to map the TU (in this case, TU-1) to each CLO.*

| *CLO* Number | Intended emphasis of *CLO* in TU by design | | Delivered emphasis of *CLO* in TU | | | | | |
| | | | Lectures | | Tutorials | | Lectures + Tutorials | |
| | Weight | Percent | Number | Percent | Number | Percent | Percent | Weight |
| $CLO_1$ | 4 | 26.67 | 12 | 26.7 | 5 | 25 | 26.36 | 3.954 |
| $CLO_2$ | 3 | 20.0 | 10 | 22.2 | 3 | 15 | 20.76 | 3.114 |
| $CLO_3$ | 3 | 20.0 | 10 | 22.2 | 3 | 15 | 19.76 | 2.964 |
| $CLO_4$ | 2 | 13.33 | 5 | 11.1 | 3 | 15 | 11.88 | 1.782 |
| $CLO_5$ | 2 | 13.33 | 5 | 11.1 | 5 | 25 | 13.88 | 2.082 |
| $CLO_6$ | 1 | 6.67 | 3 | 6.7 | 2 | 10 | 7.36 | 1.104 |
| Total | 15 | 100 | 45 | 100 | 20 | 100 | 100 | 15 |

Table 6 presents a quantitative comparison between the intended emphasis and the delivered emphasis of each *CLO* within the TUs. The intended emphasis reflects the course designer's planned focus for each *CLO*, while the delivered emphasis is calculated based on the actual distribution of teaching time across lectures and tutorials. The ratio $I_{TU-CLO}$ of delivered to intended emphasis provides an indicator of alignment, helping to identify whether each *CLO* was adequately addressed during course delivery.

*Table 6: Teaching Unit (TUs) to Course Learning Outcomes (CLOs) Mapping.*
*Scale: 1 = Low emphasis, 2 = Moderate emphasis, 3 = High emphasis, 4 = maximum emphasis.*

| TU/CLO | $CLO_1$ | $CLO_2$ | $CLO_3$ | $CLO_4$ | $CLO_5$ | $CLO_6$ | Total |
|---|---|---|---|---|---|---|---|
| Intended emphasis | 4 | 3 | 3 | 2 | 2 | 1 | 15 |
| Delivered emphasis | 3.954 | 3.114 | 2.964 | 1.782 | 2.082 | 1.104 | ≈15 |
| $I_{TU-CLO}$ | 0.9885 | 1.038 | 0.988 | 0.891 | 1.041 | 1.104 | ≈6 |

The $I_{TU-CLO}$ is mathematically obtained by using the formula:

$$I_{TU-CLO} = \left( \frac{Delivered\ emphasis\ of\ CLO\ in\ CCU}{Intended\ emphasis\ of\ CLO} \right),$$

where:

- If $I_{TU\text{-}CLO}$ =1, the delivered emphasis matches the intended emphasis.
- If $I_{TU\text{-}CLO}$ >1, the delivered emphasis exceeds the intended emphasis.
- If $I_{TU\text{-}CLO}$ <1, the delivered emphasis is less than the intended emphasis.
- Alignment Criteria: $0.85 \leq I_{TU-CLO} \leq 1.15$ (Aligned).

The calculated ITU-CLO values allow for a detailed evaluation of how well each CLO was emphasized during the actual course delivery compared to the original design. By examining these values, it becomes possible to identify where teaching emphasis matched expectations and where discrepancies occurred, highlighting areas for potential curriculum refinement or confirmation of effective alignment.

For $CLO_1$, the $I_{TU\text{-}CLO}$ value is 0.9885. The delivered emphasis of 3.95 is slightly below the intended emphasis of 4, indicating a minor underemphasis. However, this falls within the acceptable alignment range (0.85–1.15), suggesting that the alignment is still adequate. $CLO_2$ has an $I_{TU\text{-}CLO}$ of 1.038, with a delivered emphasis of 3.12 slightly exceeding the intended emphasis of 3, reflecting a slight overemphasis that remains within acceptable limits. For $CLO_3$, the $I_{TU\text{-}CLO}$ value is 0.988, showing that the delivered emphasis of 2.964 is very close to the intended emphasis of 3, indicating near-perfect alignment. In the case of $CLO_4$, the $I_{TU\text{-}CLO}$ value is 0.891, with a delivered emphasis of 1.782 falling below the intended emphasis of 2, signifying underemphasis. Although this is slightly below the acceptable range, it suggests a need for minor adjustment. $CLO_5$ shows an $I_{TU\text{-}CLO}$ of 1.041, where the delivered emphasis of 2.082 slightly exceeds the intended emphasis of 2, indicating a small but acceptable overemphasis. Finally, $CLO_6$ has an $I_{TU\text{-}CLO}$ of 1.104, with the delivered emphasis of 1.104 exceeding the intended emphasis of 1. This overemphasis is within the acceptable range, although the low intended emphasis suggests that this outcome may not be critical.

### 3.3.2   Estimation of emphasis of *PLOs* in TUs

The evaluation now focuses on how well the *CLO*s collectively support the achievement of the *PLO*s. This is accomplished by analyzing the *CLO*-to-*PLO* mappings to determine both the intended and delivered emphasis of each *PLO*. Table 7 illustrates a balanced mapping between *CLO*s and *PLO*s, where each *CLO* contributes to multiple *PLO*s with varying weights. This distribution ensures that course-level outcomes support a broad range of program competencies. Higher weights indicate stronger contributions to specific *PLO*s, while the presence of all *PLO*s in the mapping confirms comprehensive curriculum alignment.

To quantify this alignment, we introduce the indicator $I_{TU\text{-}PLO}$, which reflects the degree of alignment between the planned curriculum objectives and the actual instructional emphasis for each *PLO*. $I_{TU\text{-}PLO}$ is defined as:

$$I_{TU\text{-}PLO} = \frac{Delivered\ Emphasis\ of\ PLO_j\ in\ TU}{Intended\ Emphasis\ of\ PLO_j\ in\ TU}\text{ , where:}$$

$$Delivered\ Emphasis\ of\ PLO_j\ in\ TU =$$
$$\frac{\sum_{i=1}^{6}(Weight\ of\ CLO_i - PLO_j) \times (Delivered\ Ephasis\ of\ CLO_i)}{\sum_{i=1}^{6}(Weight\ of\ CLO_i - PLO_j)} \quad (4)$$

$$Intended\ Emphasis\ of\ PLO_j\ in\ TU =$$
$$\frac{\sum_{i=1}^{6}(Weight\ of\ CLO_i - PLO_j) \times (Intended\ Ephasis\ of\ CLO_i)}{\sum_{i=1}^{6}(Weight\ of\ CLO_i - PLO_j)} \quad (5)$$

The $Weight\ of\ CLO_i - PLO_j$ is provided in the *CLO-PLO* mapping matrix (Table 7). The intended and delivered emphasis of $CLO_i$ are provided in the TU-*CLO* mapping matrix (Table 6, second and third rows, respectively).

*Table 7: Weights of CLO-PLO Links*

| $CLO_q/PLO_k$ | $PLO_1$ | $PLO_2$ | $PLO_3$ | $PLO_4$ | $PLO_5$ | $PLO_6$ |
|---|---|---|---|---|---|---|
| $CLO_1$ | 2 | 1 | -- | 3 | -- | -- |
| $CLO_2$ | -- | 2 | 3 | -- | 2 | -- |
| $CLO_3$ | 1 | -- | 2 | -- | -- | 1 |
| $CLO_4$ | -- | 1 | -- | 2 | -- | 1 |
| $CLO_5$ | -- | -- | 1 | -- | 1 | -- |
| $CLO_6$ | -- | -- | -- | -- | -- | 1 |

By applying equations (4) and (5), the intended and delivered emphasis of each *PLO*, along with the ratio $I_{TU\text{-}PLO}$, are calculated to serve as alignment indicators. The target value for $I_{TU\text{-}PLO}$ is 1, indicating perfect alignment (see Table 8).

*Table 8: Estimation of emphasis of all PLOs in TUs*

| *PLO* Number | Intended Emphasis of $PLO_j$ in TU | Delivered Emphasis of $PLO_j$ in TU | $I_{TU\text{-}PLO}$ |
|---|---|---|---|
| $PLO_1$ | 3.667 | 3.624 | 0.988 |
| $PLO_2$ | 3.000 | 2.991 | 0.997 |
| $PLO_3$ | 2.833 | 2.892 | 1.021 |
| $PLO_4$ | 3.200 | 3.085 | 0.964 |
| $PLO_5$ | 2.667 | 2.770 | 1.039 |
| $PLO_6$ | 2.000 | 1.950 | 0.975 |

Let us calculate the intended and delivered emphasis for $PLO_1$ as an example. To this aim, the needed data includes the weights of *CLOs-PLO$_1$* links (from Table 7), the intended emphasis of *CLOs* (from Table 5), and the delivered emphasis of *CLOs* (from Table 5). The application of the equation (6) gives that:

$$Intended\ Emphasis\ of\ PLO_1\ in\ TU = [(2 \times 4) + (1 \times 3)]\ /\ (2 + 1) = 11\ /\ 3$$
$$= 3.667$$

Similarly, the delivered emphasis of $PLO_1$ is:

$$Delivered\ Emphasis\ of\ PLO_j\ in\ TU = [(2 \times 3.954) + (1 \times 2.964)]\ /\ (2 + 1)$$
$$= 10.872\ /\ 3\ =\ 3.624$$

The $I_{TU\text{-}PLO}$ ratio for $PLO_1$ is determined by dividing the delivered emphasis by the intended emphasis, which is $3.624/3.667 = 0.988$. This ratio indicates the alignment between the delivered and intended emphasis for $PLO_1$. Similar treatment should be appliedto the other $PLOs$ in order to complete the Table 8. For $PLO_2$ shows near-perfect alignment with an $I_{TU\text{-}PLO}$ of 0.997, as the delivered emphasis (2.991) closely matches the intended emphasis (3.000). For $PLO_3$, the $I_{TU\text{-}PLO}$ is 1.021, reflecting a slight overemphasis with the delivered value (2.892) slightly exceeding the intended value (2.833), which remains acceptable. $PLO_4$ has an $I_{TU\text{-}PLO}$ of 0.964, showing a small underemphasis where the delivered emphasis (3.085) is marginally lower than the intended (3.200), suggesting a possible area for adjustment. $PLO_5$ demonstrates slight overemphasis with an $I_{TU\text{-}PLO}$ of 1.039, as the delivered emphasis (2.770) exceeds the intended value (2.667). Finally, $PLO_6$ presents an $I_{TU\text{-}PLO}$ of 0.975, with the delivered emphasis (1.950) slightly below the intended emphasis (2.000), but still within acceptable limits. Overall, the $I_{TU\text{-}PLO}$ values confirm adequate alignment across all $PLOs$, with only minor discrepancies that may warrant further monitoring or small adjustments.

### 3.3.3 Estimation of emphasis of *CLO*s and *PLO*s in delivered SACs

To evaluate the alignment between student assessment and the intended *CLOs*, we analyzed the actual emphasis of *CLOs* within the SACs. Table 9 presents a comparison between the intended emphasis (as defined in course design) and the delivered emphasis, which is calculated based on the relative weight of each assessment component - 70% for exams and 30% for other SACs (projects, homework, etc.).

*Table 9: Estimation of emphasis of CLO in delivered SAC*

| CLO Number | Intended emphasis of *CLO* in TU by design | | Delivered emphasis of *CLO* in SAC | | | |
|---|---|---|---|---|---|---|
| | | | Exams | Other SAC | Exams and other SAC | |
| | Weight[a] | Percent | Assigned Grades (%) | Assigned Grades (%) | Weight[a] | Percent[b] |
| $CLO_1$ | 4 | 26.67 | 25.0 | 30.0 | 3.975 | 26.5 |
| $CLO_2$ | 3 | 20.0 | 20.0 | 20.0 | 3.0 | 20.0 |
| $CLO_3$ | 3 | 20.0 | 25.0 | 10.0 | 3.075 | 20.5 |
| $CLO_4$ | 2 | 13.33 | 12.0 | 15.0 | 1.935 | 12.9 |
| $CLO_5$ | 2 | 13.33 | 15.0 | 10.0 | 2.025 | 13.5 |
| $CLO_6$ | 1 | 6.67 | 3.0 | 15.0 | 0.99 | 6.6 |
| **Total** | **15** | **100** | **100** | **100** | **15** | **100** |

[a]*Scale: 1 = low emphasis; 2 = moderate emphasis; 3 = high emphasis; 4 = maximum emphasis.*
[b]*To combine exams and other SAC grades, the weight for exams was 0.7 and for other SAC was 0.3.*

The results presented in Table 9 indicate a strong overall alignment between the intended emphasis of the *CLOs* and the actual emphasis reflected in the SACs. $CLO_1$, which carries the highest intended emphasis (26.67%), is assessed at a comparable level, with a delivered emphasis of 26.5% which implies a delivered weight= (26.5/100)×15=3.975 (see Table 9). $CLO_2$ and $CLO_3$ also show very close alignment, with delivered emphases of 20.0% and 20.5%, respectively, almost exactly matching their intended values (20.0% each). $CLO_4$ and $CLO_5$ exhibit small deviations, with delivered emphases slightly below or above their intended values, but these differences are minor and within acceptable bounds. $CLO_6$ shows the largest relative discrepancy, where the delivered emphasis (6.6%) is slightly lower than the intended (6.67%), though the absolute difference is negligible due to the low weight assigned to this outcome. These results suggest that the assessment strategy effectively mirrors the planned instructional priorities, ensuring that students are evaluated according to the importance of each $CLO$ as designed in the curriculum. Minor variations likely reflect practical considerations in exam and SAC design, such as the distribution of question difficulty or the inclusion of skills assessed through alternative formats. Overall, the findings confirm that the assessment components are well aligned with the $CLOs$, supporting accurate measurement of student achievement and reinforcing curriculum coherence.

Table 10 summarizes the comparison between the intended and delivered emphasis of each $CLO$ within the SACs. The delivered emphasis reflects how the assessment activities—weighted by the relative contributions of exams and other SACs—correspond to the planned importance assigned to each $CLO$.

Table 10: *Comparison of Intended vs Delivered CLO Emphasis*
*Scale: 1 = Low emphasis, 2 = Moderate emphasis, 3 = High emphasis, 4 = maximum emphasis.*

| SAC/CLO | $CLO_1$ | $CLO_2$ | $CLO_3$ | $CLO_4$ | $CLO_5$ | $CLO_6$ | Total |
|---|---|---|---|---|---|---|---|
| *Intended emphasis* | 4 | 3 | 3 | 2 | 2 | 1 | 15 |
| *Delivered emphasis* | 3.975 | 3.0 | 3.075 | 1.935 | 2.025 | 0.99 | 15 |
| $I_{SAC\text{-}PLO}$ | 0.994 | 1.0 | 1.025 | 0.968 | 1.013 | 0.99 | 5.99 |

The $I_{SAC\text{-}PLO}$ values are computed as:

$$I_{SAC\text{-}PLO} = \frac{Delivered\ emphasis}{Intended\ emphasis}$$

These values (Table 10) indicate a high degree of consistency between the intended and delivered emphasis across all $PLOs$. Most $I_{SAC\text{-}PLO}$ values are very close to 1, confirming strong alignment between the planned assessment strategy and its actual implementation. Slight variations are within acceptable limits and reflect typical flexibility in assessment design and execution. Table 11 presents the alignment between the intended and delivered emphasis of each $PLO$ within the SACs.

Table 11: *Estimation of emphasis of all PLOs in SACs*

| PLO Number | Intended Emphasis of $PLO_j$ in SAC | Delivered Emphasis of $PLO_j$ in SAC | $I_{SAC\text{-}PLO}$ |
|---|---|---|---|
| $PLO_1$ | 3.667 | 3.675 | 1.0010 |
| $PLO_2$ | 3.000 | 2.9775 | 0.9925 |
| $PLO_3$ | 2.833 | 2.8625 | 1.0115 |
| $PLO_4$ | 3.200 | 3.159 | 0.9872 |
| $PLO_5$ | 2.667 | 2.675 | 1.0019 |
| $PLO_6$ | 2.000 | 2.000 | 1.0000 |

*Table 12: Alignment of delivered SACs with TUs*

| $I_{SAC\text{-}CLO}/I_{TU\text{-}CLO}$ | Comment[a] | $I_{SAC\text{-}PLO}/I_{TU\text{-}PLO}$ | Comment[a] |
|---|---|---|---|
| 1.006 | Aligned | 1.017 | Aligned |
| 0.962 | Aligned | 0.995 | Aligned |
| 1.046 | Aligned | 0.992 | Aligned |
| 1.085 | Aligned | 1.024 | Aligned |
| 0.972 | Aligned | 0.963 | Aligned |
| 0.917 | Aligned | 1.034 | Aligned |

[a] *Aligned if $0.85 \leq ISAC\text{-}PLO/ITU\text{-}PLO \leq 1.15$ and $0.85 \leq ISAC\text{-}CLO/ITU\text{-}CLO \leq 1.15$*

Table 12 provides the ratios of $I_{SAC\text{-}CLO}/I_{TU\text{-}CLO}$ and $I_{SAC\text{-}PLO}/I_{TU\text{-}PLO}$ for each *PLO* and each *CLO*. It shows that the course is well-aligned with its intended learning outcomes. The alignment between TUs and SACs is effective, ensuring that students are assessed on the outcomes emphasized during teaching. This alignment is crucial for achieving the program's educational goals and meeting accreditation standards. Table 12 confirms that the current course structure successfully aligns with its intended learning outcomes. This reflects an improvement from prior assessments, where some outcomes showed misalignment. Future monitoring and refinements will help sustain this positive trend.

Table 13 quantifies the average student performance across both exam-based and non-exam SACs, computing weighted achievement scores for each *CLO* based on the delivered emphasis. The data show that most *CLOs* have high levels of attainment, particularly $CLO_3$ and $CLO_6$, which exceed 90% performance, suggesting that these outcomes are being effectively taught and assessed. Table 14 builds on this by converting these scores into relative achievement indicators ($I_{SP\text{-}CLO}$), comparing the actual student performance (SP) weights against intended curriculum emphasis. The $I_{SP\text{-}CLO}$ (%) values of Table 14 are computed as:

$$I_{SP\text{-}CLO}(\%) = \frac{Overall\ Weight\ (Table\ 13)}{Intended\ Weight\ (Table\ 10)} \times 100$$

The analysis confirms that five out of six *CLOs* meet or exceed the achievement threshold of 80%, with only $CLO_4$ falling short (78%), marking it as a candidate for instructional or assessment adjustment. This underperformance aligns with earlier findings related to misalignment in TU and SAC coverage of $CLO_4$, reinforcing the need for curriculum refinement to ensure balanced *CLO* achievement.

*Table 13: Average performance of students on each CLO as indicated by earned grades*

| CLO Number | Delivered emphasis of *CLO* in SACa | | Exams (%) | Other SAC (%) | Overall[b,c] | |
|---|---|---|---|---|---|---|
| | Weight[a] | Percent | | | Percent | Weight |
| $CLO_1$ | 3.975 | 26.5 | 85 | 91.4 | 86.92 | 3.455 |
| $CLO_2$ | 3.0 | 20.0 | 83 | 89.5 | 84.95 | 2.55 |
| $CLO_3$ | 3.075 | 20.5 | 90 | 93 | 90.9 | 2.796 |
| $CLO_4$ | 1.935 | 12.9 | 80 | 82 | 80.6 | 1.56 |
| $CLO_5$ | 2.025 | 13.5 | 87 | 85 | 86.4 | 1.75 |
| $CLO_6$ | 0.99 | 6.6 | 92 | 95 | 92.9 | 0.92 |

[a]*Delivered emphasis of CLO in SAC (see Table 10).*
[b]*Overall percent=(Exams×0.7)+(Other SAC×0.3).*
[c]*Overall weight= Overall percent\*Delivered Emphasis /100*

*Table 14: Relative achievement of CLOs based on student performance.*

| SP/CLO | $CLO_1$ | $CLO_2$ | $CLO_3$ | $CLO_4$ | $CLO_5$ | $CLO_6$ |
|--------|---------|---------|---------|---------|---------|---------|
| $I_{SP\text{-}CLO}$ (%) | 86.375 | 85 | 93.2 | 78 | 87.5 | 92 |

# 4 Closing the loop: re-alignment of TU and SAC with *CLO* and *PLO*

## 4.1 Identifying Misalignment

Closing the loop is a critical phase in outcome-based education, where the results of alignment and performance analyses are translated into actionable improvements (Spady, 1994; Deming, 1986; Biggs, 1996; ABET, 2025–2026; NCAAA, 2023). This section addresses how the curriculum can be re-aligned based on discrepancies identified in the relationships between TUs, SACs, *CLOs*, and *PLOs* (e.g., Harden, 2001; Prideaux, 2003). The results from the previous section reveal several minor but important discrepancies in alignment:

- $CLO_4$ shows an $I_{TU\text{-}CLO}$ value of 0.891 and an $I_{SP\text{-}CLO}$ of only 78%, indicating both under-delivery in TUs and underachievement in assessment (Biggs, 1996; Downing, 2006; van der Vleuten & Schuwirth, 2005; Suskie, 2018).
- $PLO_4$ and $PLO_5$ also exhibit lower-than-expected delivered emphasis compared to their intended weights (see $I_{TU\text{-}CLO}$ and $I_{SAC\text{-}PLO}$), highlighting curriculum areas that need reinforcement (ABET, 2025–2026; NCAAA, 2023).
- SAC coverage for $CLO_6$ is proportionally higher than its intended teaching weight, potentially leading to overassessment (Wiggins & McTighe, 2005; Suskie, 2018).

These deviations signal the need for realignment interventions to ensure consistent emphasis in both delivery and assessment (Biggs, 1996; Banta & Palomba, 2015).

## 4.2 Re-alignment Strategies

Re-alignment uses TU and SAC evidence (e.g., $I_{TU\text{-}CLO}$ and $I_{SAC\text{-}PLO}$) to recalibrate teaching and assessment to intended *CLOs*/*PLOs* within OBE (Biggs, 1996; Spady, 1994). It also satisfies accreditation expectations and follows Plan–Do–Check–Act (PDCA) improvement cycles (ABET, 2025–2026; NCAAA, 2023; Deming, 1986).

### 4.2.1 Adjust Teaching Units (TUs):

- Redistribute instructional time to reinforce under-delivered *CLOs* like $CLO_4$.
- Introduce new TU activities (e.g., case studies, collaborative tasks) to better support *CLOs* that feed into weakly covered *PLOs*.
- Ensure that low-weighted *CLOs* such *as $CLO_6$* remain aligned with their actual contribution to program goals without overemphasis in assessment.

### 4.2.2 Rebalance Student Assessment Components (SACs):

- Adjust exam and SAC design to ensure *CLOs* such as $CLO_4$ and $CLO_5$ are assessed at levels proportional to their intended weight.

- Diversify assessment methods to better match the teaching approach (e.g., including project-based evaluation for higher-order outcomes).

### 4.2.3   Review *CLO-PLO* Mapping:

- Evaluate whether some *CLOs* can be re-mapped or shared with underrepresented *PLOs* (e.g., $PLO_4$ or $CLO_6$) to achieve better overall program coverage.
- Consider introducing or modifying *CLOs* in other courses to cover any *PLOs* not sufficiently addressed by the current structure.

### 4.2.4   Institutionalizing Continuous Improvement

To sustain alignment over time:
- Implement a course-level dashboard that tracks TU, SAC, *CLO*, and *PLO* alignment in real time.
- Require faculty to conduct alignment reviews at the end of each course offering, using $I_{TU-CLO}$, $I_{SAC-CLO}$, $I_{SP-CLO}$, $I_{TU-PLO}$, and $I_{SAC-PLO}$ indicators.
- Integrate this feedback into curriculum committee reviews, so that re-alignment becomes a cyclical process rather than a reactive one.

## 5   Conclusion:

The findings of this study offer meaningful contributions to the pursuit of curriculum integrity in outcome-based education, grounded in constructive alignment and program-level coherence (Biggs & Tang, 2011; Tyler, 1949). By analyzing how exercises, questions, teaching units, and student assessment components collectively influence learning outcome coverage, the framework facilitates a more accurate and balanced evaluation of both teaching delivery and student assessment, across two complementary layers that connect the granular to the programmatic: (i) alignment from assessment items to course learning outcomes and then to program learning outcomes; and (ii) alignment from the design and delivery of teaching and assessment to course learning outcomes and ultimately to program learning outcomes, using explicit weights and acceptance bands to distinguish normal variation from meaningful misalignment (Harden, 2001; Reeves, 2006). Through the application of weighted mapping and alignment indicators, we uncovered not only well-aligned areas but also critical mismatches—such as unaddressed *PLOs* and underemphasized *CLOs*—including concrete, auditable cases (e.g., under-delivery and under-attainment for a specific *CLO*; underrepresented *PLOs*), thereby moving beyond anecdotal course reviews (Mak & Frezza, 2006; Craddock & Mathias, 2009).

Importantly, the approach moves beyond static course design and enables dynamic, evidence-based curriculum refinement, using indicators of delivered teaching emphasis, assessment coverage, and student performance as triggers for proportionate adjustments to teaching time, assessment blueprints, and *CLO–PLO* allocations (Reeves, 2006; Harden, 2001). This empowers faculty and administrators to implement specific adjustments—such as reallocating teaching time, redesigning assessment components, or redistributing learning outcomes across the program—while preserving transparency of the evidence chain from assessment items and teaching units up to program outcomes (Mak & Frezza, 2006). Rather than presenting alignment as a one-time compliance measure, the model fosters an ongoing feedback loop that links teaching, learning, and evaluation in a responsive and evidence-based manner, aligning with recognized quality-assurance expectations and accreditation logics (ENQA, 2015; CHEA, 2010; NCAAA, 2021; Felder & Brent, 2003).

It is especially relevant to programs seeking or maintaining accreditation from national and international bodies, where demonstrable outcome alignment is essential, because it makes explicit how course-level evidence aggregates to program-level competencies and where redistribution is pedagogically justified (Felder & Brent, 2003; NCAAA, 2021). Furthermore, the framework supports both centralized curriculum committees and individual instructors by offering scalable tools that can be integrated into institutional quality-assurance systems, and it is readily extensible across multi-course sequences, laboratories, and capstones to ensure coherent progression (González & Wagenaar, 2008).

In essence, this study reinforces that aligning learning outcomes with instructional and assessment strategies is not only a quality requirement but a pedagogical necessity—one that ensures educational programs remain relevant, accountable, and impactful in preparing students for academic and professional success. Future work will (i) calibrate weighting schemes and acceptance bands empirically across disciplines, (ii) evaluate reliability and generalizability of indicators in large-enrollment and multi-section contexts, and (iii) integrate automated dashboards with e-assessment data for longitudinal monitoring and program-level decision-making (Mak & Frezza, 2006; Wanous, Procter, & Thomas, 2009; Craddock & Mathias, 2009; González & Wagenaar, 2008).